The Importance of Funding Space-Based Research

Devan Taylor

devantaylorpublications@gmail.com

Author Note

Dedicated to David Andrew Taylor.

## Abstract

When it comes to conversations about funding, the question of whether the United States should be spending its resources on space-based research often rears its head. Opponents of the idea tend to share the opinion that the resources would be better spent helping citizens on the ground. With an estimated homeless population around 562,000 throughout the country [1], roughly 39.4 million Americans (12.3% of the population) living below the poverty level [2], and 63.1 million tons of food waste per year [3], it's hard to argue that the United States does not have its share of problems that need to be addressed. However, a history of space-based research has proven time and time again to bring forth advances in technology and scientific understanding that benefit humans across the globe and provide crucial protection for life on Earth.

*Keywords*:  space, research, united states, nasa

## NASA's Budget

Before the argument can be made for continuing to fund space-based research, it's important to understand NASA's (the leader in United States-based spaceflight technologies) current budget. NASA (National Aeronautics and Space Administration) reported $22,559,000,000 (twenty-two billion five hundred fifty-nine million dollars) for their total budget in 2020 and has requested a budget of $25,246,000,000 (twenty-five billion two hundred forty-six million dollars) for 2021, an increase of $2,687,000,000 (two billion six hundred eighty-seven million dollars) [4]. The estimated outlays for the United States government's budget in 2021 are $4,829,000,000,000 (four trillion eight hundred twenty-nine billion dollars) [5]. With a comparison of these numbers, NASA's requested budget for 2021 equates to roughly 0.52%, or about one half of one percent, of the total estimated spending for the United States' government in 2021. If the United States government's spending equated to $1.00, NASA's budget would be equivalent to less than a single penny.

**Protecting Life on Earth**

Despite their low budget compared to other government funded programs, NASA takes on one of the most important tasks to protect life on Earth: monitoring and researching methods to deflect asteroids. Asteroids are by far one of the largest threats faced by humanity. A single impact event of sufficient size could disrupt global climates, cause the deaths and/or relocation of millions of humans, and even make Earth completely uninhabitable.

Currently slated to launch in late July of 2021, NASA's DART (Double Asteroid Redirection Test) mission is designed to test the capabilities of asteroid deflecting technology [6]. The DART mission will include sending a spacecraft to (65803) Didymos, a binary near-Earth asteroid system, and deliberately crashing it into the moonlet of the system to measure the amount of deflection achieved. The moonlet (the smaller body of the Didymos two body system) is typical of the size of asteroids that pose a significant threat to Earth, and as such will provide NASA with valuable data they can use in the future when the need to deflect an asteroid arises.

If caught early enough, only a small deflection is needed to change a hazardous object's course from impacting Earth. The earlier the deflection, the less deflection is needed. Conversely, the less notice there is, the harder it is to deflect hazardous objects. Because of this, asteroid monitoring systems and missions like NASA's DART are extremely important in the fight to protect Earth from the only natural disaster humans can defend against; potentially saving countless lives.

In the words of "your personal astrophysicist," Neil deGrasse Tyson, "Asteroids have us in their sight. The dinosaurs didn't have a space program, so they're not here to talk about this problem. We are, and we have the power to do something about it. I don't want to be the embarrassment of the galaxy, to have had the power to deflect an asteroid, and then not, and end up going extinct" [7]. The technology needed to monitor, categorize, and deflect potentially harmful near-Earth objects depends solely on having enough funds to do so. When the risks are as detrimental as complete extinction of humans, and possibly all surface-dwelling animal life, it's hard to argue that the effort and money is better spent somewhere else. Independent organizations, such as the Planetary Society, rely on funding from the general public to spread awareness of planetary defense and attempt to educate the U.S. Congress of its importance.

### A Better Understanding of Home

A plethora of NASA missions orbit Earth, giving humans across the globe valuable information guiding the general public's day to day life. The Atmospheric Infrared Sounder (AIRS), onboard NASA's Aqua satellite, is used by weather prediction centers across the globe by creating 3-dimensional maps of cloud properties and land and air temperatures. "The forecast improvement accomplishment alone makes the AIRS project well worth the American taxpayers' investment." -A quote from Dr. Mary Cleave of NASA's Science Mission Directorate in a news release from 2005 [8]. Anyone that has used weather forecasts to guide their decisions may have data from the AIRS mission to thank, and it isn't the only instrument in orbit providing valuable data for weather forecasts.

AIRS is one of only six instruments onboard the Aqua satellite, itself one of many satellites monitoring Earth's weather and atmosphere jointly referred to as the Earth Observing System (EOS). Other EOS satellites measure Earth's interior, gravity, magnetic field (which plays an important role in protecting Earth from harmful solar radiation), carbon cycle, biogeochemistry, atmospheric composition, water cycles, ocean topography, and even climate change [9]. Not only do these satellites give humans a better understanding of how Earth's various systems function, but they also allow humans to make predictions for how climate change will affect future generations and how to prevent further damage. This data is of immense value to the global population, not just citizens of the United States. Continued funding is required to make sure humans stay up to date with the latest technology and reach the best understanding possible of the planet they call home.

## Technological Advances

Known colloquially as "NASA Spinoff Technology," many gadgets enjoyed by the general public, or used by emergency responders and medical professionals, have direct roots in NASA spaceflight missions. Researchers at NASA's Jet Propulsion Laboratory (JPL) revolutionized imaging technology in the 1990s by way of the Complementary Metal-Oxide-Semiconductor (CMOS) image sensor. These sensors would later be used in almost all modern digital cameras, including smartphones [10]. "CMOS imagers offer significant advantages in terms of lower power, low voltage, flexibility, cost and miniaturization. These features make them very suitable especially for security and medical applications" [11]. When it comes to imaging devices onboard spacecrafts, low power requirements, low cost, and miniaturization are all advantageous properties. Artificial retinas, or small ocular implants used to stimulate neural cells in the retina, take advantage of CMOS technology to return partial vision to blind patients [11].

The Global Positioning System (GPS) is used daily by many all over the country, whether it's to explore new areas on vacation, to travel to unknown locations for their career, or just to check out the new ice cream shop across town. The GPS is a radionavigation system based on a constellation of satellites in orbit operated by the United States Air Force (USAF). NASA works closely with the USAF to improve GPS technology as it has valuable use in space-based operations [12]. Proper funding is essential for the continued improvement and application of the GPS for use by the general public and spaceflight missions alike.

NASA's Spinoff publication highlights the various gadgets that make use of technology created by or directly influenced by NASA research, all of which have only come to fruition because of the funding of space-based research. Other technologies directly inspired by NASA research include: The Ejenta system that allows doctors to monitor patients remotely [13], inexpensive ventilators made from non-traditional parts (as not to compete for supplies with other ventilator manufacturers) to aid patients during the COVID-19 pandemic [14], remote sensing technology to help fire fighters track the paths of forest fires [15], satellite imagery to help keep track of freshwater usage for crops [16], the Hazard Analysis and Critical Control Point (HACCP) system for reducing foodborne illness that keeps foods in supermarkets safe for consumers [17], the tunable laser spectrometer (TLS) for detecting methane gas [18], Psionic lidar scanners used in self-driving cars [19], and impact absorbing foam used in aircraft seats to protect passengers as well as improved bra designs for a more comfortable fit [20]. More examples of how NASA technology has been used and transformed to adapt to the needs of people all over the world for use in public safety, environmental protection, consumer products, transportation, information technology, and healthcare can be found online at spinoff.nasa.gov. Without the funding of NASA, many gadgets used to save lives and create a better day to day living for people across the globe would not exist in their current state.

## Countdown to Extinction

Even if all Earth-based obstacles faced by humans are overcome; harmful asteroids are deflected, climate change is controlled, population growth is controlled, and humans do not succumb to the threats of nuclear annihilation or global pandemics, humans still have a time limit for existence on the planet. This is because the life-giving star in the center of the solar system, the Sun, is a ticking time bomb.

Although the Sun is currently in a stable part of its development, converting hydrogen into helium within its core via nuclear fusion and balancing the inward pull of gravity, all stars have a life cycle and eventually die off. When the Sun runs out of hydrogen to convert into helium in its core, the inward pull of gravity will start to take over, shrinking the Sun's core until it compresses the gases enough to begin burning hydrogen in a shell around the core. This will transition the Sun from a yellow dwarf star into a red giant. During this transition, the shell around the core will increase in size significantly, overtaking the orbit of Mars and heating Earth to a much higher degree than it currently does. This overheating will cause the oceans to boil and evaporate, surely not a comfortable day for any lifeforms still inhabiting the planet. As the water from the oceans is split into hydrogen and oxygen by cosmic rays in the upper atmosphere, Earth will lose all its water to space, leaving it a barren, dry, rock orbiting the now much larger Sun [21].

If it wasn't apparent already, humans will not be able to survive on an Earth devoid of water with blisteringly high temperatures. Before this point, it will be necessary for humans to set up colonies on planets around other stars if they wish to overcome annihilation. This will be the most challenging task ever taken on by humans. It will require extensive advances in space-faring technology and worldwide cooperation between nations that can only be accomplished by the continued funding and advancement of space-based research. Without space-based research, the human race lives with a countdown timer to extinction.